\documentstyle[prl,aps,epsfig,multicol]{revtex} 
\begin{document} 
\title{The Hertz contact in chain elastic collisions} 
 
\author{P. Patr\'\i cio} 
\address{ 
Instituto Superior de Engenharia de Lisboa\\ 
Rua Conselheiro Em\'\i dio Navarro 1, P-1949-014 Lisboa, Portugal
} 
 
\date{January 2004} 
 
\maketitle 
 
\begin{abstract} 
 
A theoretical analysis about the influence of the Hertz elastic 
contact on a three body chain collision is presented. In spite of 
the elastic character of the collision, the final velocity of 
each particles depends on the particular interaction between them. A 
system involving two elastic spheres falling together one in top 
of the other under the action of gravity, and colliding with an 
horizontal hard wall is studied in detail. The effect of the Hertz 
contact interaction can be easily put in evidence for some particular situations. 
 
\end{abstract} 

\vspace{2mm}
 
\begin{multicols}{2} 
 
\section{Introduction} 
 
When two balls are dropped together, 
both vertically aligned and hitting the ground at the same time, 
an interesting rebound effect is obtained. If the smaller ball
is above, it may rise surprisingly high into the air. This effect may still be enhanced 
if one drops a chain of balls, ordered by their relative weight, 
the lightest ball at the top. 
 
This system was already discussed in pedagogical articles more than 30 years ago, 
assuming perfect and independent elastic collisions between each pair of balls 
\cite{Mellen,Harter,Kerwin,Huebner,Mellen2}. It was considered an attractive 
and ideal system to study elastic collisions. 
However, if the balls are dropped together, 
all collisions will happen at the same time. Because they are not instantaneous, 
the final velocity distribution of the balls will be dependent on the particular 
interactions between them, even in those cases where the interactions are elastic, 
{\sl i.e.}, with no loss of energy. 
In their beautiful article \cite{Harter}, the class of W. G. Harter considered 
at some point non-instantaneous collisions. To study them, they introduced a simple 
quasi-linear model, which could be valid if we think not of spheres but of 
uni-dimensional springs. The dynamical differential equations were solved 
using an analog computer program. However, they concluded that the quasi-linear 
model did not agree with their experimental results, and they proposed 
a phenomenological non linear model. 
The interaction field was then obtained directly from empirical data, 
by dropping a ball onto a painted flat metal surface from varying heights. 
In this way the interaction could be plotted as a function of the spot size, 
and eventually as a function of the depression of the ball. 
Although the strategy to obtain the non linear interaction was remarkable, 
the discussion that followed was not sufficiently developed. 
In fact, from the theoretical point of view and as a first approximation, 
the analysis of the non-instantaneous multiple collisions effect 
should take into account the interaction between 
two elastic spheres obeying Hooke's law, which is always valid for 
sufficiently small deformations. This problem is known in the literature 
as the Hertz contact \cite{Landau}. 
 
In this article, a system involving multiple collisions between 
two different elastic spheres, one on top of the other, with the 
same initial velocity, hitting a planar hard horizontal wall is 
considered. In Section~II, the results obtained for instantaneous 
elastic collisions are briefly reviewed. Next, the Hertz contact 
is introduced in the mathematical model of the system. At the end, 
the results of this theoretical analysis are presented. The main 
differences between independent and non-instantaneous collisions 
are outlined and discussed. 
 
\section{Independent collisions} 
 
Consider a system with two elastic spheres, one on top of the other, 
falling with the same velocity $v$, colliding with the ground, as represented 
in Fig.~1. 
 
\begin{figure}[htb] 
\par\columnwidth=20.5pc 
\hsize\columnwidth\global\linewidth\columnwidth 
\displaywidth\columnwidth 
\centerline{\epsfxsize=100pt\epsfbox{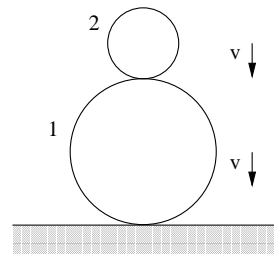}} 
\caption 
{Two different elastic spheres falling with the same velocity 
colliding with the ground.} 
\label{fig:1} 
\end{figure} 
 
Assume a very small gap between the spheres, such that all collisions 
can be treated individually. 
First, the lower sphere hits the rigid wall and returns with the same 
velocity $v$. Next, the spheres collide elastically. 
The final velocities $u_1$ and $u_2$ (of lower and upper sphere respectively) 
can be calculated through momentum and energy conservation principle, 
and are independent of the particular interaction. 
If $m=m_2/m_1$ is the spheres mass ratio, we can write: 
\begin{eqnarray} 
\label{eq:1} 
&u_1=\frac{1-3m}{1+m}v 
 \\ 
\label{eq:2} 
&u_2=\frac{3-m}{1+m}v, 
\end{eqnarray} 
where the positive sign corresponds to the upward orientation. 
The upper sphere may rebound with a velocity up to three times greater 
than its initial one, for a vanishingly small mass ratio. 
That is, if its mass is much smaller than the mass of the lower sphere. 
 
In fact, the rebound may be enhanced if we add to this system 
extra spheres, all falling with the same initial velocity $v$. 
To study this effect, consider first the elastic collision between 
two spheres with a mass ratio $m$ but with different initial velocities. 
The upper sphere still moves with a downward velocity $v$. 
The lower sphere has now previously collide with another sphere and moves upward 
with a velocity $av$ ($a>1$). The final velocities are now given by: 
\begin{eqnarray} 
\label{eq:3} 
&u_1=\frac{a-(a+2)m}{1+m}v 
 \\ 
\label{eq:4} 
&u_2=\frac{1+2a-m}{1+m}v. 
\end{eqnarray} 
Naturally, equations (\ref{eq:1}) and (\ref{eq:2}) are recovered if $a=1$.
For a system with $n$ spheres, the upper one may achieve a 
final velocity of $(2^n-1)v$, if all mass ratios are vanishingly small. 
 
Another interesting limit to this system, for a certain 
distribution of mass ratios, concerns the case for which all lower 
spheres stop, and only the upper one rises into the air. In this 
situation, all initial energy was transmitted to only one sphere. 
In the simple system composed of only two spheres, the lower one 
has null final velocity for $m=1/3$, whereas the upper sphere 
rebounds to reach twice its initial velocity. 
 
\section{Non instantaneous collisions} 
 
However, if the spheres fall together, the assumption of 
independent collisions may no longer be accurate. For the system 
represented in Fig.~1, both collisions - of the lower sphere with 
the ground and of the two spheres - will happen at the same time. 
To obtain the final velocities, after the spheres separation, it 
is important to know the particular interaction between them. In 
the following subsections, one presents the simplest possible 
interaction: it will be assumed that all solids respect Hooke's 
law of elasticity, valid for sufficiently small deformations. The 
dynamical equations for this model will then be established and 
solved. 
 
\subsection{The Hertz contact} 
 
The theory of the elastic contact between solids was first 
studied by H. Hertz, and it can be followed in detail in Ref.~\cite{Landau}. 
For two isotropic and homogeneous elastic spheres, 
with radius of curvature respectively of $R_1$ and $R_2$ (see Fig.~2), 
the elastic potential energy $E_p$ depends on the combined deformation $h$ as: 
\begin{equation} 
\label{eq:5} 
E_p=\frac{2}{5}er^\frac{1}{2}h^\frac{5}{2}, 
\end{equation} 
where the reduced radius $r=R_1R_2/(R_1+R_2)$ and the reduced 
elastic constant $e$ is dependent on the Young moduli $E_1$, $E_2$ 
and on the Poisson coefficients $\sigma_1$, $\sigma_2$. It is 
given by: 
\begin{equation} 
\label{eq:6} 
e=\frac{4}{3}\left(\frac{1-\sigma_1^2}{E_1}+\frac{1-\sigma_2^2}{E_2}\right)^{-1}. 
\end{equation} 
The Young modulus increases with the solid rigidity, whereas the 
Poisson coefficient is typically slightly smaller than $1/2$. 
 
\begin{figure}[htb] 
\par\columnwidth=20.5pc 
\hsize\columnwidth\global\linewidth\columnwidth 
\displaywidth\columnwidth 
\centerline{\epsfxsize=100pt\epsfbox{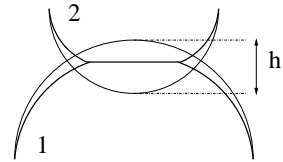}} 
\caption 
{Elastic contact between two spheres.} 
\label{fig:2} 
\end{figure} 
 
The elastic energy~(\ref{eq:5}) is valid for a static deformation. 
Nevertheless, it can be considered a good approximation for the 
situation studied in this article if the velocity $v$ is much 
smaller than the sound velocities of the solids involved. It is 
interesting to notice that this approximation fits well the 
experimental potential energy obtained by the class of W. G. 
Harter~\cite{Harter}, in the first regime of small ball 
depressions. For large depressions, the energy increased with a 
greater power of $h$, which is in agreement with the fact that 
Hooke's elastic regime only take into account the first terms of 
the elastic energy power expansion on the deformation. 
 
It is instructive to calculate the maximum height of deformation 
and the contact time of two colliding spheres. Suppose $\nu$ is 
the relative velocity of the spheres before the collision, and 
$\mu=m_1m_2/(m_1+m_2)$ their reduced mass. Then, the energy 
conservation principle reads: 
\begin{equation} 
\label{eq:7} 
\frac{1}{2}\mu\left(\frac{dh}{dt}\right)^2+ 
\frac{2}{5}er^\frac{1}{2}h^\frac{5}{2}=\frac{1}{2}\mu\nu^2. 
\end{equation} 
The maximum height of deformation occurs when the relative 
velocity is zero and can be written: 
\begin{equation} 
\label{eq:8} 
h_M=\left(\frac{5\mu}{4er^\frac{1}{2}}\right)^\frac{2}{5}\nu^\frac{4}{5}. 
\end{equation} 
The collision time corresponds to the time in which the 
deformation goes from $0$ to $h_M$ and back to $0$ again. It is 
given by: 
\begin{equation} 
\label{eq:9} 
\tau=2\left(\frac{25\mu^2}{16e^2r\nu}\right)^\frac{1}{5} 
\int_0^1\frac{dx}{\sqrt{1-x^\frac{5}{2}}}\approx 
3.21\left(\frac{\mu^2}{e^2r\nu}\right)^\frac{1}{5}. 
\end{equation} 
These are the relevant length and time for elastic collision 
problems, and they will be useful to write down dimensionless 
dynamical equations. 
 
\subsection{Dynamical equations} 
 
For the system represented in Fig.~1, the total potential energy 
is given by the sum: 
\begin{equation} 
\label{eq:10} 
E_p=\frac{2}{5}e_{01}r_{01}^\frac{1}{2}h_{01}^\frac{5}{2}+ 
\frac{2}{5}e_{12}r_{12}^\frac{1}{2}h_{12}^\frac{5}{2}, 
\end{equation} 
where the first term takes into account the interaction between the first sphere 
and the rigid ground, with infinite Young modulus and radius of curvature, 
and the second term is the interaction between both spheres. 
The deformations depend on the positions $x_1$ and $x_2$ of the centers 
of the spheres (the ground defining the reference position): 
\begin{eqnarray} 
\label{eq:11} 
&h_{01}=H(R_1-x_1), 
 \\ 
\label{eq:12} 
&h_{12}=H\left((R_1+R_2)-(x_2-x_1)\right). 
\end{eqnarray} 
The function $H(x)=x$ if $x>0$ and $H(x)=0$ otherwise. 
The equations of motion 
$m_i\ddot{x}_i=-\partial E_p/\partial x_i$, ($i=1,2$), 
will simply be written: 
\begin{eqnarray} 
\label{eq:13} 
&m_1\ddot{x}_1=e_{01}r_{01}^\frac{1}{2}h_{01}^\frac{3}{2}- 
e_{12}r_{12}^\frac{1}{2}h_{12}^\frac{3}{2}, 
 \\ 
\label{eq:14} 
&m_2\ddot{x}_2=e_{12}r_{12}^\frac{1}{2}h_{12}^\frac{3}{2}. 
\end{eqnarray} 
If one introduces natural units, $L=(m_1v^2/e_{01}r_{01}^\frac{1}{2})^\frac{2}{5}$
and $T=L/v$, which are associated with
the maximum height of depression and the time of contact 
between the ground and the first sphere, 
it is possible to define dimensionless variables. 
The equations of motion may be written in a simpler form: 
\begin{eqnarray} 
\label{eq:15} 
&\ddot{x}_1=h_{01}^\frac{3}{2}-kh_{12}^\frac{3}{2}, 
 \\ 
\label{eq:16} &m\ddot{x}_2=kh_{12}^\frac{3}{2}. 
\end{eqnarray} 
They only depend on two parameters, 
\begin{eqnarray} 
\label{eq:161} 
&m=m_2/m_1,
 \\ 
\label{eq:162} 
&k=e_{12}r_{12}^\frac{1}{2}/e_{01}r_{01}^\frac{1}{2}. 
\end{eqnarray} 
Note that $k$ is defined as a ratio of the reduced elastic constants and
radius. If the infinite rigidity of the planar wall is taken into account,
we have $e_{01}>e_{12}$ and $r_{01}>r_{12}$. 
For this system, one concludes that $k<1$.

\subsection{Numerical resolution} 
 
\begin{figure}[htb] 
\par\columnwidth=20.5pc 
\hsize\columnwidth\global\linewidth\columnwidth 
\displaywidth\columnwidth 
\centerline{\epsfxsize=220pt\epsfbox{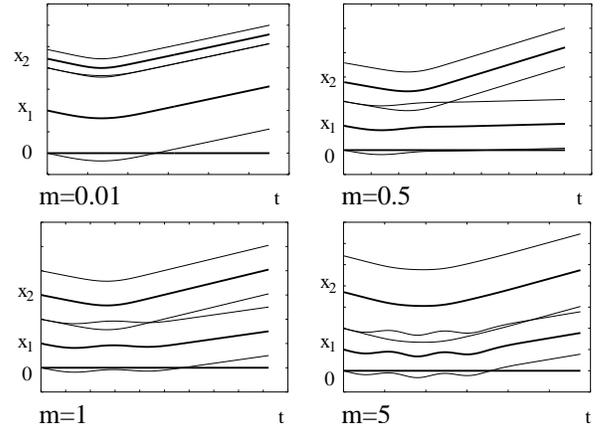}} \caption { The 
positions $x_1$ and $x_2$ as a function of time, for spheres
with the same Young modulus and density, and four different mass ratios 
$m=0.01, 0.5, 1, 5$. The axis have arbitrary units. The bold lines 
represent the ground and the center of the spheres. In thinner 
lines, $x_1\pm R_1$ and $x_2\pm R_2$ are also represented, to indicate in 
which periods of time the interaction exists.} \label{fig:3} 
\end{figure}

The set of differential dynamical equations
(\ref{eq:15}) and (\ref{eq:16}) was solved numerically using a simple 
and standard Euler method, for different values of the parameters. 
Fig.~\ref{fig:3} shows several solutions found for the positions 
$x_1$ and $x_2$ as functions of time, starting when the two 
spheres initially in contact hit the ground with the same velocity 
$v$, until they separate with constant final velocities $u_1$ and 
$u_2$. The axis have arbitrary units. 
The solutions were 
calculated for spheres with the same Young modulus and density,
but four different mass ratios $m=0.01, 0.5, 1, 5$ 
(wich were obtained choosing appropriate sphere radii). 

Surprisingly, for a very small mass ratio, the spheres have 
almost the same velocity $u_1\approx u_2\approx v$ after the 
rebound. This situation  contrasts with the results obtained 
considering independent collisions, for which the maximum velocity 
gain would be expected. 
In fact, if the mass ratio $m$ is negligible, it is possible to
see from Eq.~\ref{eq:16} that the deformation $h_{12}$ will
also be very small, exherting almost no influence in the motion of the
lower sphere (see Eq.~\ref{eq:15}). The latter sphere rebounds with the wall
and returns with velocity $u_1\approx v$.
Since $h_{12}\approx R_2\approx 0$, it may be concluded that $R_1\approx x_2-x_1$.
Then $\ddot{x}_1\approx\ddot{x}_2$, 
which means that both spheres stick together during the
collision as if they were one whole body \cite{Silva}.

It is also interesting to notice that for 
mass ratios greater than one (in the case of the figure with $m=5$) the 
three body collision gets more complex, with the bottom sphere 
making several rebounds between the ground and the other sphere. 
 
It is important to mention that the final velocities do not depend 
on the amplitude of the spheres' deformations during the 
collisions. If both spheres were more rigid, but keeping the same 
ratio $k$, the deformations could be very small, and still the 
final velocities would be equal to the ones represented in 
Fig.~\ref{fig:3}. 

\begin{figure}[htb] 
\par\columnwidth=20.5pc 
\hsize\columnwidth\global\linewidth\columnwidth 
\displaywidth\columnwidth 
\centerline{\epsfxsize=220pt\epsfbox{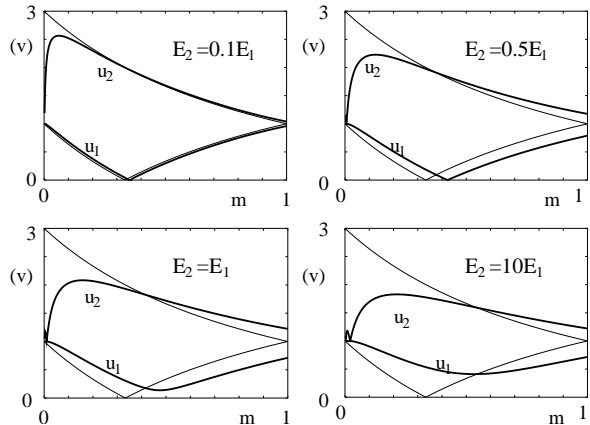}} \caption {Velocity 
distributions $u_1$ and $u_2$ as functions of $m<1$. The thin 
lines correspond to the results obtained for independent 
collisions and they are independent of the ratio between
the Young moduli. In bold, the 
velocities obtained considering the Hertz contact for 
$E_2/E_1=0.1,0.5,1,10$.} \label{fig:4} 
\end{figure}

In Fig.~\ref{fig:4}, the bold lines represent the final velocity 
distributions $u_1$ and $u_2$ as functions of $m<1$, for different 
values of $E_2/E_1=0.1,0.5,1,10$. The thin lines show the results of 
equations (\ref{eq:1}) and (\ref{eq:2}), considering independent 
collisions.  

The independent collisions results are recovered for small
values of the ratio $E_2/E_1$, and consequently for small $k$. 
This is particularly true for the final velocity distribution of the lower sphere $u_1$. 
In fact, if $k\rightarrow 0$, it is possible to see 
from Eq.~\ref{eq:15} that in a first moment,
the motion of the lower sphere
will depend mainly on its interaction with the wall --
the deformation $h_{12}$ between both spheres
will have a negligible contribution 
when it is compared with the deformation
$h_{01}$ between the lower sphere and the wall.
However, if the collision time between both spheres is large enough
(as it is shown in Eq.~\ref{eq:9},
the collision time increases for small $k$),
after the lower sphere-wall interaction,
the terms containing the deformation $h_{12}$ will be the only ones
present in the equations of motion.
Therefore, the collisions can be considered as almost independent and subsequential. 
Nevertheless, as the final velocity $u_2$ is concerned, 
the convergence does not occur for all values of $m$:
it is possible to see in Fig.~\ref{fig:4} ($E_2/E_1=0.1$)
that the final speed $u_2\rightarrow v$ as the mass ratio $m\rightarrow 0$.

As $E_2/E_1$ (or $k$) increases, 
small changes can be seen in the final velocity distributions,
and the independent collisions approximation is no longer appliable.
These changes are limited 
because the infinite rigidity assumption requires $k<1$.
In fact, the velocity distributions for $E_2/E_1\gg 1$
is not very different from the one represented for $E_2/E_1=10$

\begin{figure}[htb] 
\par\columnwidth=20.5pc 
\hsize\columnwidth\global\linewidth\columnwidth 
\displaywidth\columnwidth 
\centerline{\epsfxsize=220pt\epsfbox{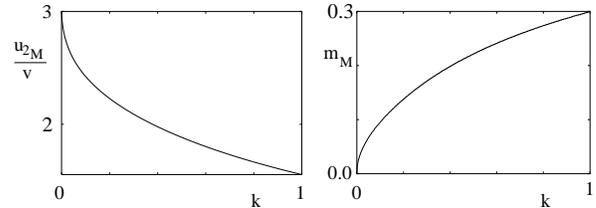}} \caption {
Maximum rebound speed $u_{2_M}$ and corresponding mass ratio $m_M$ as
a function of $k<1$
(according to the infinite rigidity of the wall assumption).
} \label{fig:5} 
\end{figure}

The left part of Fig.~\ref{fig:5} represent 
the rebound maximum speed $u_{2_M}$ as a function of $k<1$.
On the right,
the mass ratio $m_M$ corresponding to the maximum rebound speed is also plotted.
$m_M$ decreases rapidly to zero as $k\rightarrow 0$.

The discontinuity on the slope observed 
for the final velocity $u_1$ in Fig.~\ref{fig:4} ($E_2/E_1=0.1$) 
corresponds to the critical value 
$m_c$ for which the lower sphere stops completely. If $m>m_c$, 
this sphere rebounds twice off the ground, 
before it reaches the final velocity represented in the figure. 
If the ratio between the Young moduli increases, the discontinuity changes first
its position ($E_2/E_1=0.5$),
and eventually it disappears ($E_2/E_1=1$ and larger values).
 
\begin{figure}[htb] 
\par\columnwidth=20.5pc 
\hsize\columnwidth\global\linewidth\columnwidth 
\displaywidth\columnwidth 
\centerline{\epsfxsize=120pt\epsfbox{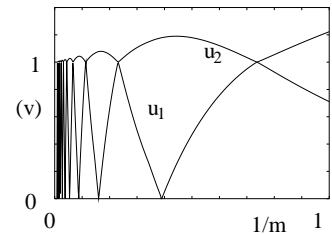}} \caption {Velocity 
distributions $u_1$ and $u_2$ as functions of $1/m<1$, for 
$E_2/E_1=1$. These results correspond to the results obtained only 
considering non-instantaneous collisions.} \label{fig:6} 
\end{figure}

However, these discontinuities will happen again for larger values of $m$, 
as the lower sphere rebounds more and more in between the ground 
and the upper sphere. This curious feature can be seen in 
Fig.~\ref{fig:6}, which represents the velocity distributions 
$u_1$ and $u_2$ as functions of $1/m<1$, for 
$E_2/E_1=1$. As $1/m$ approaches $0$, {\sl i.e.}, when the lower sphere becomes
much lighter than the top one, the number of collisions 
increases to infinity. 
 
\section{Discussion and conclusions} 

In this article, the influence of the Hertz contact on multiple 
chain collisions was studied in detail. It was shown that the 
independent collisions approximation is no longer valid for small 
values of the mass ratio $m$ and large values of the rigidity ratio $k$. 

To understand the results obtained here, it is important to
consider the typical two-body times of collision (see Eq.~\ref{eq:9}).
First, suppose $m$ is of the order of unity.
If $k$ is small, the two-body collision time between the
lower sphere and the wall, $\tau_{01}$, is smaller than the two-body collision time between
both spheres, $\tau_{12}$. 
As $k\rightarrow 0$, the first collision
occurs almost instantaneously, and subsequently, the two spheres collide.
Thus, the results obtained considering this limit approach the independent collisions results.

However, as $k$ increases, $\tau_{01}$ also increases. 
The collisions become more evolved, and it is difficult to treat them separately. 

Despite its complexity, the final velocity distributions $u_1$ and $u_2$
approach each other for large $k$ and small $m$.
In this limit ($m\rightarrow 0$), as it was prooved through the analysis
of the equations of motion, both spheres behave as if they belonged to the same body,
changing their velocities together as they interact with the ground.
At the end, both spheres rebound with equal final velocities $u_1\approx u_2\approx v$. 

The results obtained with this
model were summarized in Fig.~\ref{fig:5}, which clearly indicates
the best parameters $m$ and $k$ to experimentally put in evidence
the influence of the Hertz contact on chain elastic collisions.

{\bf Acknowledgments} 
 
I would like to thank J. P. Silva for getting me 
interested in the multiple chain collision problem, for very 
useful discussions, and also for a critical reading of the 
article. I also would like to thank J. M. Tavares for his 
enthusiasm and for enlightening comments. 

I also acknowledge the Centro de F\'\i sica Te\'orica
e Computacional da Universidade de Lisboa for its facilities.

\end{multicols} 
 
\end{document}